\RequirePackage{lineno}
\documentclass[aps,preprint,showpacs,superscriptaddress,groupedaddress]{revtex4} 
\usepackage{bm}
\usepackage{amsmath}
\usepackage{amsfonts}%
\usepackage{amssymb}%
\usepackage{graphicx}
\usepackage{color}
\usepackage{hyperref}
\usepackage{verbatim} 
\usepackage{caption}

\begin{document}
\newcommand{\AR}{|\Delta D/\Delta f|_0}  
\newcommand{\ARN}{A_r}  
\newcommand{\brp}{{\bf r}^{\prime}}  
\newcommand{\bq}{{\bf q}}
\newcommand{\bs}{{\bf s}}  
\newcommand{\bsp}{{\bf s}^{\prime}} 
\newcommand{\zp}{z^{\prime}}
\newcommand{\br}{{\bf r}}     
\newcommand{\brc}{{\bf r}_0}
\newcommand{\gammaref}{\Gamma_{\mathrm{ref}}}
\newcommand{\omegaref}{\omega_{\mathrm{ref}}}
\newcommand{\homegaref}{\hat{\omega}_{\mathrm{ref}}}
\newcommand{\fref}{\tilde{f}_{\mathrm{ref}}}
\newcommand{\rt}{\tilde{R}}
\newcommand{\yt}{\tilde{y}}
\newcommand{\homega}{\hat{\omega}}
\newcommand{\meter}{\mathrm{m}}
\newcommand{\nm}{\mathrm{nm}}
\newcommand{\cpar}{\mathrm{C}}
\newcommand{\mdel}{m_{\delta}}
\newcommand{\hk}{\tilde{k}} 
\newcommand{\gam}{\dot\gamma}
\newcommand{\ndelta}{N_{\delta}}
\newcommand{\dgam}{\hat{\dot\gamma}}
\newcommand{\ar}{\mathrm{A}_r}
\newcommand{\hx}{\hat{{\bf x}}}
\newcommand{\hz}{\hat{{\bf z}}}
\newcommand{\bn}{\hat{{\bf n}}}
\newcommand{\hv}{\hat{v}}
\newcommand{\ha}{\hat{\ddot x}}
\newcommand{\ot}{\omega t}
\newcommand{\dom}{\Delta \omega}
\newcommand{\dof}{\Delta f}
\newcommand{\hs}{\hat{\sigma}}
\newcommand{\hsa}{\hat{\sigma}^{(a)}}
\newcommand{\hsm}{\hat{\sigma}_m}
\newcommand{\ii}{\mathrm{i}}
\newcommand{\hZ}{\hat{\mathcal{Z}}}  
\newcommand{\impre}{\mathcal{Z}_{R}}  
\newcommand{\impim}{\mathcal{Z}_{I}}  
\newcommand{\imp}{\mathrm{Z}}  
\newcommand{\impff}{\mathcal{Z}_{ff}}  
\newcommand{\impnf}{\mathcal{Z}_{nf}}  
\newcommand{\impa}{\mathcal{Z}^{(a)}}  
\newcommand{\impf}{\mathcal{Z}^{(0)}}  
\newcommand{\impfree}{\mathcal{Z}^{(\mathrm{L})}}  
\newcommand{\impaf}{\mathcal{Z}^{(a_\mathrm{fluid})}}  
\newcommand{\impdna}{\mathcal{Z}^{(\mathrm{DNA})}}  
\newcommand{\impldna}{\mathcal{Z}^{(\mathrm{L-DNA})}}  
\newcommand{\vb}{v^{(0)}} 
\newcommand{\ypar}{d} 
\newcommand{\rp}{{\bf r}_c} 
\newcommand{\fext}{{\bf F}^{\mathrm{ext}}} 
\newcommand{\ffluid}{{\bf F}^{\mathrm{drag}}} 
\newcommand{\vf}{{\bf v}} 
\newcommand{\vs}{{\bf v}_{S}} 
\newcommand{\vp}{{\bf v}_p} 
\newcommand{\vres}{\mathrm{v}_0} 
\newcommand{\bu}{{\bf u}} 
\newcommand{\bv}{{\bf v}} 
\newcommand{\bvp}{{\bf v}^{\prime}} 
\newcommand{\vpx}{v^{\prime}_x} 
\newcommand{\bx}{\hat{{\bf x}}} 
\newcommand{\by}{\hat{{\bf y}}} 
\newcommand{\re}{\mathrm{Re}}
\newcommand{\im}{\mathrm{Im}}
\newcommand{\bG}{\boldsymbol{\mathrm{G}}}
\newcommand{\hbG}{\boldsymbol{\mathrm{\hat{G}}}}
\newcommand{\hG}{\mathrm{\hat{G}}}
\newcommand{\bsig}{\boldsymbol{\sigma}} 
\newcommand{\bsigtot}{\boldsymbol{\Pi}} 
\newcommand{\bGc}{\boldsymbol{\mathrm{G}}_0}
\newcommand{\bsind}{\boldsymbol{\mathcal{S}}^{\mathrm{ind}}} 
\newcommand{\sindx}{{\mathcal{S}}^{\mathrm{ind}}_x} 
\newcommand{\bfin}{{\bf F}^{\mathrm{ind}}} 
\newcommand{\finx}{F^{\mathrm{ind}}_x} 
\newcommand{\str}{{\bf S}^{\mathrm{ind}}} 
\newcommand{\av}{\overline{{\bf v}}^{(s)}}
\newcommand{\avs}{\overline{{\bf v}_{s}}^{(v)}}
\newcommand{\avsx}{\overline{v_{s}}^{(v)}}
\newcommand{\avp}{\overline{{\bf v}_{p}}^{(s)}}
\newcommand{\basw}{\overline{\boldsymbol{\sigma}_{\mathrm{wall}}}}
\newcommand{\asw}{\overline{\sigma_{\mathrm{wall}}}}
\newcommand{\mqcm}{m_{\mathrm{QCM}}}


\bibliographystyle{unsrt} 

\title{Quantitative reinterpretation of quartz-crystal-microbalance experiments
  with adsorbed particles using analytic hydrodynamics}
\author{Marc Melendez Schofield}
\affiliation{Departmento de Fisica de la Materia Condensada, Universidad Autonoma de Madrid,
and Institute for Condensed Matter Physics, IFIMAC.
Campus de Cantoblanco, Madrid 28049, Spain}
\author{Rafael Delgado Buscalioni}
\email[]{rafael.delgado@uam.es}
\affiliation{Departmento de Fisica de la Materia Condensada, Universidad Autonoma de Madrid,
and Institute for Condensed Matter Physics, IFIMAC.
Campus de Cantoblanco, Madrid 28049, Spain}
\begin{abstract}
    Despite being a  fundamental tool in soft  matter research, quartz crystal microbalance (QCM) analyses  of discrete macromolecules in liquid so far  lack a firm theoretical  basis. Currently, acoustic signals  are  qualitatively  interpreted using  ad-hoc  frameworks based  on effective  electrical  circuits,  effective springs  and trapped-solvent       models      with       abundant      fitting parameters. Nevertheless, due to  its extreme sensitivity, the QCM technique pledges  to become  an accurate predictive  tool.  Using unsteady   low  Reynolds   hydrodynamics   we  derive   analytical
    expressions  for  the  acoustic  impedance  of  adsorbed  discrete spheres.   Our   theory  is  successfully  validated   against  3D simulations and  a plethora of experimental  results covering more than a decade of research on proteins, viruses, liposomes, massive nanoparticles,  with  sizes  ranging   from  few  to  hundreds  of nanometers.   The excellent  agreement  without fitting  constants clearly indicates that  the acoustic response is  dominated by the hydrodynamic   impedance,   thus,    deciphering   the   secondary contribution  of  physico-chemical  forces will  first  require  a
    hydrodynamic-reinterpretation of QCM.
\end{abstract}
\maketitle

\section{Introduction}

The quartz  crystal microbalance (QCM) is one  of the  most
versatile  tools to study subtle effects  in soft  matter,
resolving forces in the pico-Newton to nano-Newton range
and  nano-gram masses. Due to its  low  operating  cost, sensor  compactness,  real-time  data, label-free operation
and subnanogram sensitivity, QCM has become a fundamental tool in  analytical chemistry and biophysics research.
The number of applications (from nanotribology to health care, environmental monitoring \cite{Krim2012,Johannsmann2015,Bragazzi2015,Rodahl1997} and even crude oil \cite{Balestrin2019}) is huge and cannot be exhaustively listed here.
QCM has also become   one  of  the   important  techniques in
biosensing for DNA \cite{Liu2004,Nie2007,Electra2020,prapp2020} and other biomolecules \cite{Ruzgas2019} including virus detection \cite{virus07}. These distinct features make QCM competitive with other common analytical and detection tools \cite{Fogel2016} such as optical DNA  detection via fluorescence-labeled oligonucleotides,
surface  plasmon resonance  \cite{Hui2014}, or electrochemical assays. 
Subtle nanometric phenomena such as variations in contact forces, molecular stiffness \cite{Busscher2017,Olson2012stiff}, kinetics   of   adsorption or  bio-molecular interactions \cite{2013biointeract} are routinely {\em sensed} using QCM.
However, in these liquid environments, QCM lacks the theoretical foundation required to become a {\em measurement technique}.

The  idea of  using the  inverse piezoelectric  effect to sense mass,  which is in essence the QCM, was born for experiments in a vacuum.  The surface of a cut of crystal  quartz exposed to an AC potential, oscillates at MHz
 frequency  and the  inertia of tiny  amounts of  deposited material
creates a detectable reduction in the  oscillation frequency $\Delta f<0$.
Sauerbrey \cite{Saue} converted this phenomenon into a mass balance, by showing
the proportionality between $-\Delta f$ and  the  deposited  mass per  unit  surface $\mqcm$,
\begin{linenomath*}
 \begin{equation}
   \label{eq:mqcm}
   \mqcm= -\mathrm{C}  \Delta f/n
 \end{equation}
\end{linenomath*}
 where $n$ is the overtone of the surface wave (odd integer $n\leq 13$).
 The mass sensitivity constant, typically $\mathrm{C=17.7\ \mathrm{ng} \cdot\mathrm{cm}^{-2}\mathrm{Hz}^{-1}}$,
 reveals an extremely small limit of detection, as $\Delta  f  \sim -0.1\  \mathrm{Hz}$ represents $1.7\ \mathrm{ng/cm^2}$.
 
 Interpreting QCM in liquids faced challenges, many of them still unresolved.
 In a liquid, viscous forces propagate the wall oscillation
 upwards, moving a layer of fluid of about 3 times the so-called penetration  depth
 $\delta=(2\eta/\rho\omega)^{1/2}$ (here  $\omega=2 \pi f$,  $\eta$ is
 the fluid  viscosity and  $\rho$ its density). In  water, $\delta\propto \eta^{-1/2}$
 typically ranges  from 71 to 238 nm. The resulting laminar flow, called Stokes
flow, creates wall viscous stress oscillating with a $45^o$ phase lag
with respect the surface  motion. The out-of-phase component damps the wall motion.
Its decay rate $2 \pi \Gamma$ is directly measured in ``ring-down'' sensors (QCM-D) \cite{Johannsmann2015} while in forced QCM,
this dissipative effect broadens the spectra, with half bandwidth $\Gamma$
and quality factor $f/(2\Gamma)$. The new actor, ``dissipation'' $D=2\Gamma/f$, introduces another channel of information in liquids. In Newtonian fluids \cite{kanazawa1985,Ricco1987} $-\Delta  f$ and $\Delta \Gamma$ are equal and proportional to the mass of moving fluid. Viscoelastic films  \cite{Johannsmann1992,Voinova1999} present different contributions which can be traced using 1D laminar flow equations.
However, QCM was soon used to investigate all sorts of
soft discrete 3D objects, for which an analytical approach has so far been elusive. The QCM   technique faced proteins
\cite{mateos2016monitoring,Milioni2017},   DNA   strands   \cite{tsortos2008quantitative}, supported      lipid     bilayers  \cite{Mazur-lipids-qcm2017,VitaalaSLB2018,Richter2005}, polymers \cite{Marx2003}, vesicles \cite{Keller1998},  liposomes  \cite{reviakine2012,Electra2020,prapp2020},  viruses \cite{Tellechea2009,virus07},   different kinds   of   nano and  microparticles \cite{Pomorska2010,Liu2004,Nie2007,Ruzgas2019},  bacteria   \cite{Busscher2017},  living  cells \cite{Braunhut2005}, crude  oil \cite{Balestrin2019}  and more.

Experiments urgently required ways to rationalize 
 the distinct acoustic features and
 peculiar behaviors of these discrete analytes.
The adsorbed mass predicted from $\Delta f$ using Eq. \ref{eq:mqcm}
was seen to significantly differ (usually appearing larger) than
other independent measurements of $m$,
e.g. using scanning electron microscopy (SEM) \cite{Bingen2008,Grunewald2015}.
 For more than one decade such a difference, measured
 by the mass ratio $H=1-m/\mqcm$ \cite{Bingen2008,Johannsmann2008a,Sadowska2020},
 has been explained using the ``trapped solvent'' model \cite{Bingen2008}
 which assumes that the extra QCM mass is due to solvent molecules being
 trapped by the analyte and moving concomitantly with it.
Despite the reported deficiencies \cite{Grunewald2015}, several versions of this
model are still routinely used to interpret experiments \cite{Sadowska2020}.

Another unexplained puzzling phenomenon concerns the frequency inversion.
As the analyte size (or QCM frequency) is increased, $\Delta  f$ becomes more and more
negative until above a certain size (or frequency) it suddenly 
becomes positive \cite{Melendez2020}. Phenomenological models
were designed to reproduce such behavior.
The coupled-resonator model \cite{dybwad85}
is based on a series of masses connected with effective springs 
representing analyte-wall contacts \cite{Olsson2011,dybwad85,Tarnapolsky2018,Johannsmann2015} placed either  in  parallel  (Kelvin-Voight) or  in  series  (Maxwell model) \cite{Busscher2017}. This model predicts a transition from  ``inertial'' ($\Delta  f<0$) to ``elastic'' ($\Delta f>0$) response at high frequency, when the large contact
stiffness overpowers the inertia of deposited mass.  
Imaginary springs are also added to act as dampers, introducing the
concept  of ``viscous  load''  ($\Delta \Gamma  >0$)  of the adsorbed
structure. Tuning the model parameters permits fitting experimental data and gauging
different analyte ``stiffnesses'',  adsorbed ``mass''  or analyte-wall ``interactions''. However, the coupled-resonator model completely neglects the role of
the solvent hydrodynamics. These phenomenological
pictures  very  much  constitute  the  basis  of  present
analyses  \cite{Olson2012stiff,Busscher2017}. Quoting  Tarnapolsky and
Freger  \cite{Tarnapolsky2018},  QCM-D  has ``mainly  become  a
comparative tool  in particle  adhesion research.  Unfortunately, such
development  lacks an  adequate  quantitative  model''.

About one decade  ago, simulations started to  highlight the relevance
of        hydrodynamics         in        discrete-particle        QCM
\cite{Tellechea2009,reviakine2011hearing}. Coverage  effects
such as the decrease of  the acoustic ratio $-\Delta D/\Delta  f$ with $\Delta f$
were    qualitatively     reproduced in     2D    simulations
\cite{Tellechea2009,reviakine2011hearing} and later in 3D \cite{Johannsmann2015a,Gillissen2018}, revealing a hydrodynamic origin, which has not yet been
theoretically explained. The relevance of the
particle shape  \cite{Gillissen2017a} was also analyzed.
Recently, it was proved that hydrodynamics lie behind the extreme sensitivity of QCM to how broadly mass is distributed over
the resonator \cite{prapp2020} and also that it is responsible for
anti-Sauerbrey responses  ($\Delta f>0$) \cite{Melendez2020}.

Before introducing the concept of hydrodynamic impedance,
a comment on the phasor formalism is in order. The resonator position
can be  expressed  as $\tilde{x}(t) =  \mathrm{Re}[x\, \exp(-\ii \hat{\omega}  t)]$
where $\hat{\omega} \equiv 2\pi f - \ii 2\pi \Gamma$  is the complex frequency
and $x$ is its phasor (note that $\Gamma >0$ implies an exponential decay). This complex number
determines its phase lag with respect some time
reference. The central phasor quantity in QCM is the impedance $\imp=\asw/\vres$
which, following the small  load approximation ($\Delta f/f  << 1$),
relates the overall tangential wall-stress $\asw\equiv\hat{\bf x} \cdot  \basw\cdot  \hat{\bf z}$ with the complex frequency shift \cite{Johannsmann2015},
\begin{linenomath*}
  \begin{equation}
    \label{eq:sla}
  \Delta f - \ii \Delta \Gamma = \ii f \frac{\imp}{\pi \imp_Q},
  \end{equation}
\end{linenomath*}
  where the impedance of the quartz crystal cut is usually
  $\imp_Q=8.8\times 10^{6}\ \mathrm{kg/(m^2 s)}$.

  The origin of the hydrodynamic impedance is simple \cite{prapp2020,Melendez2020}:
  any force acting on the analyte propagates fluid momentum to the resonator,
  creating extra wall-stress which is measured by the QCM. It is important to note that particle-forces
  arise not only from molecular linkers, adhesion forces, etc.,
but they are {\em also induced by the fluid traction itself}. In fact,
we shall show that the QCM response is dominated by
{\em fluid-induced forces}. In any case, to understand the QCM response
one needs to determine the lag-time required to transmit the
analyte-force to the wall. This time
crucially depends on the vertical coordinate $z$
because (as shown below) the fluid-momentum propagator is proportional to $\exp(-\alpha z)$, with  $\alpha  = (1-\ii)/\delta$.
Without loss of generality, let the wall phasor $x=x_0$ be a real number and consider an oscillatory force in phase with the wall ($F$ is real),
acting at some point located at $z=d$. This force
transfers a hydrodynamic stress $(F/A) \exp(-\alpha d)$ to the wall
($A$  is  the   resonator  area) creating an impedance $Z= [F/(A \omega x_0)] \ii \exp[-\alpha d]$.  If the force is placed at the wall
($d=0$) this leads to $\mathrm{Re}[\imp]=0$ and $\mathrm{Im}[\imp]>0$
(or, from Eq. \ref{eq:sla}, $\Delta  \Gamma =0$ and $\Delta f <0$).
In the QCM jargon this would correspond to an inertial load. But the very same
force applied at $d=(\pi/2) \delta$ would then be understood as a
purely viscous load ($\Delta \Gamma >0$ and $\Delta f =0$), while farther away it would become an elastic load ($\Delta  f >0$). This simple example clearly illustrates
 the need for a rigorous hydrodynamic reinterpretation of QCM  signals.

 In  general, the values of
 $\Delta  f$  and $\Delta \Gamma$ result from summing up
 the propagation of all forces acting on each point of the {\em ensemble} of
 analytes. This leads to a far-from-trivial convolution
 expression, which should be derived using zero-Reynolds unsteady hydrodynamics \cite{Kim,Pozrikidis2016,Felderhof2012}. Indeed, the hydrodynamics of QCM gathers all the  difficulties one  might expect:
 the semi-bounded unsteady flow lacks spherical symmetry
 and obtaining the perturbative flow created by the particle
 (which creates the extra wall stress)
 requires solving the dynamics of the analyte, which, in turn,
 is coupled to the {\em fluid-induced} forces.
 While such  an intertwined problem can be partially
 tackled in the case of point particles \cite{morrison2018},
 many QCM analytes (liposomes, nanoparticles) are far from being ``points''
 and reach the size of the penetration depth $R\sim \delta \sim 100 \mathrm{nm}$.
Fortunately, QCM senses the total stress over the surface which
simplifies the analytical expressions for the impedance of finite adsorbed particles, derived below. Comparison with 3D simulations and abundant available experimental data proves that our approach
is valid up to $R/\delta < 2$.
Notably, although we just consider free particles (wall-particle forces are absent)
the theory shows an excellent agreement with quite a disparate set of
experiments, without any fitting constants.
This result urgently calls for a quantitative reinterpretation of QCM signals
starting from the dominant role of hydrodynamics, adding to the predictive power of QCM and becoming a tool for {\em measuring} relevant forces, due to molecular/structural elasticity, adhesion, ionic-strength or other long-ranged physico-chemical interactions with the substrate.

  

\section{Results}

\subsection{Theory}
  
We consider a sphere of radius $R$ and density $\rho_p$ whose center, located at $\rp =(0,0,\ypar)$, is at distance $\ypar$ from the QCM plane $z=0$.
The QCM resonator oscillates at angular frequency $\omega$
in the $x$ direction with velocity $\vres\,\cos(\omega t)$ and its
amplitude $x_0$ is small enough (typically around $2 \mathrm{nm}$) to neglect non-linear couplings.
The total impedance $\imp$ sums up all the forces (per area) acting on the surface.
As customary, the baseline is set at the impedance of
the base Stokes flow (equal to $\eta \alpha$)
so we consider stress in excess of that reference.
The forces acting on the wall are either directly due to the particle
(impedance noted as $\imp_{\mathrm{pw}}$) or
to the fluid ({\em hydrodynamic impedance}, $\imp_{\mathrm{hydro}}$),
\begin{linenomath*}
\begin{equation}
  \imp= \imp_{\mathrm{pw}} + \imp_{\mathrm{hydro}}
\end{equation}
\end{linenomath*}
In turn, $\imp_{\mathrm{pw}}$ has contributions from the particle inertia
and from wall-particle forces (adhesion, molecular linkers, etc.). The latter
will not be considered hereafter, so as to isolate
the hydrodynamic effects. The particle inertia is just the Archimedean force due to
the acceleration of the excess particle mass so that $\imp_{\mathrm{pw}}=\ii n m_e \omega$.
Implicitly, we have assumed that the adsorbed particle velocity $u$ concomitantly  follows
that of the resonator $u=\vres$.
Here $n$ is the particle's surface density and $m_e= (\rho_p-\rho) V_p$
is the excess in mass with respect to the displaced fluid
($V_p=4\pi R^3 /3$ is the particle volume).  In terms of the scaled impedance
$\hZ_{\mathrm{pw}}\equiv \imp_{\mathrm{pw}}/(6\pi \eta n R)= -(2/9) (\rho_e/\rho) (\alpha R)^2$, with $\rho_e = \rho_p-\rho$.
This is precisely the Sauerbrey contribution to the impedance,
with zero dissipation and negative frequency shift 
(i.e, $\mathrm{Re}[Z_{\mathrm{pw}}]=0$  and   $\mathrm{Im}[Z_{\mathrm{pw}}]>0$, as $\mathrm{Im}[\alpha^2]<0$).

{\em Any} force acting on the particle is transferred back to the fluid (Newton's third law) as a force density field which propagates momentum to the surface
and creates extra wall-stress (detected by the QCM device as frequency $\Delta f$ and dissipation $\Delta D$ shifts). As stated, here we will only consider {\em fluid-induced forces}.
The fluid velocity field can be expressed
as $\vf={\bf v}_{\infty} +\vp$, where $\vp$ is the perturbative
flow created by the particle presence and the ambient flow $v_{\infty}$
is here ascribed to the base laminar
Stokes profile ${\bf v}_{\infty}=\vs=v_s(z) {\bx}$. Its
phasor satisfies $v_s^{\prime \prime} -\alpha^2 v_s =0$
(prime denotes spatial derivation) with boundary conditions $v_s(0) = \vres$
and $v_s(\infty)=0$. The solution, $v_s(z) = \vres \, \exp(-\alpha z)$, unveils the exponential propagator of momentum mentioned above.

The hydrodynamic  impedance requires evaluating the tangential  stress due to the perturbative  flow at the
resonator $z=0$. Such a flow is governed by the Green   function   tensor  field   $\bG(\br,\brp)$  of the problem (Methods). For instance,
a point-particle at $\rp$ receiving  an oscillatory force (phasor)
$-{\bf  F}$ creates a flow field $\bv_p(\br) = \bG(\br,\rp) {\bf F}$.
A finite  particle propagates the forces  acting on each differential
element $dS^{\prime}$ on its surface, which (in the absence of
wall or external forces) is induced  by the  fluid
pressure at the particle surface, so
\begin{linenomath*}
\begin{equation}
  \label{eq:pex}
  \bv_p(\br) = \oint \bG(\br,\brp) \bsigtot(\brp) \cdot \bn dS^{\prime}.
\end{equation}
\end{linenomath*}
Here $\bsigtot(\brp)$  is the local  fluid pressure tensor,  $\bn$ is
the outwards  surface vector and  the integral runs over the particle
surface with $dS^{\prime}$ centered at $\brp$.
As $\vs(z=0)=\vres$, one  has to impose $\bv_p=0$ at $z=0$
and  at $z\rightarrow \infty$;  these  boundary  conditions  are
inherited by $\bG(\br,\brp)$ (see Methods). In the present setup, however, an explicit derivation of $\bv_p$  faces serious  difficulties.
The Fax\'en theorem route  consists in integrating the  no-slip condition
$\bv=\bu$ at the particle surface to impose a translational (and in
general  rotational)  constraint  $\av \equiv  \oint \bv  dr^2/(4 \pi
R^2)=\bu$  onto Eq.  \ref{eq:pex}. Providing $\bu$ should lead to $\bsigtot$. In general, though, $\bu$ for suspended particles has to be determined from the particle  equation of  motion
(for a free particle, $-\ii  m_p \omega  \bu =  \oint \bsigtot(\br)\cdot \bn  dS$).
Due  to the  lack of  spherical symmetry
this route becomes impracticable and, to complicate matters further, in this setup  $\bG(\br,\brp)$ has no closed analytical form\cite{Pozrikidis2016}.  A  second route, based on hydrodynamic reflections \cite{Kim,morrison2018}
is to expand  $\bsigtot$ into ambient and  perturbative parts
$\bsigtot    =\bsigtot_S   +\bsigtot_p$. Introducing this form into
Eq.\ref{eq:pex}  leads  to  a series expansion with operators acting on
$\bv_s$ and involving increasing powers of $\bG$. But again,
this requires a closed  form for $\bG$ in real space.
Fortunately, Felderhof \cite{Felderhof2012} demonstrated that it  is possible to derive the Fourier transform
of  $\bG(\br,\brp)$ in  the  xy-plane which, as we will show shortly,
suffices  for  our  purposes.  The pressure tensor
has a viscous stress $\bsig$ and a kinetic pressure contribution
which create a viscous $Z_v$ and kinetic $Z_k$ impedance derived below. The kinetic stress is just the virial pressure
created by fluid inertial forces relative to the base flow $\ii \omega (\rho_p \bu -\rho \avs)$ (with $\avs = [3/(4\pi R^3)] \int \bv(\br) d^3 r$ the average fluid velocity over the particle volume).
The viscous stress includes a dominant contribution
from the Stokes base flow $\bsig_S$ and another
from the perturbative flow $\bsigtot_p$.
The excess pressure tensor at the particle surface can thus be written as
\footnote{We note that inserting
\ref{eq:bsigtot} into Eq. \ref{eq:pex} leads to an expression for the perturbative flow similar to that obtained from the reciprocal theorem in the case of neutrally buoyant particles \cite{Kim,Pozrikidis2016}. Also, Eq. \ref{eq:bsigtot} should include an entropic contribution (order $k_BT$)
due to the particle Brownian motion \cite{usabiaga2013minimal,Delong2014,atzberger2011jcp}. However, this contribution can be neglected due to the extremely fast QCM oscillation frequency \cite{prapp2020}.}, \setcounter{footnote}{16}
\begin{linenomath*}
\begin{equation}
  \label{eq:bsigtot}
  \bsigtot(\brp) = \bsig_S(\brp) + \ii \omega \left(\rho_p \bu -\rho \avs \right) {\brp} + \bsigtot_p,
\end{equation}
\end{linenomath*}
where $\bsig_S= \eta v_S^{\prime}(z) \hat{\bf x} \hat{\bf z}$
and $\bsigtot_p$ is expected to be small
for $R/\delta < O(1)$ and shall be neglected in this analysis.
This approximation finds support later in the comparison to simulations and experimental results.
We assume that the particle moves in the $x$ direction
$\bu = u \hat{\bf x}$ and note that $\avs=\avsx \hat{{\bf x}}$.

To evaluate the net shear stress at the wall,
one needs to integrate over the resonator plane ($z=0$),
the tangential stress $\eta \partial_z \bv_p\cdot \hx$
due to the perturbative flow in Eq. \ref{eq:pex},
\begin{linenomath*}
\begin{equation}
  \asw= n \eta \int_{z=0} dS \, \oint_{r=a} \hx\cdot \left[\partial_z \bG(\br,\brp)\right]_{z=0} \bsigtot\cdot{\hat{\bf n}}\, dS^{\prime}    
\end{equation}
\end{linenomath*}
Owing to the planar symmetry of the system \cite{Felderhof2012}
$\bG(\br,\brp)=\bG(\bs-\bs^{\prime}; z,z^{\prime})$
where $\br = \bs + z\hat{\bf k}$ and $\bs$ lies on the xy-plane.
This permits the introduction of the Fourier transform on the xy-plane,
\begin{linenomath*}
$$
\bG(\bs -\bs^{\prime},z,z^{\prime}) = \int d\bq e^{\ii \bq \cdot({\bs}-{\bsp})} \hat{\bG}(\bq,z,z^{\prime}) 
$$
\end{linenomath*}
to obtain
\begin{linenomath*}
\begin{equation} 
  \asw = \\n \eta \int_{z=0} d^2{\bf s} \oint_{r=a} dS^{\prime}    \int d \bq e^{\ii \bq \cdot({\bs}-{\bsp})} \hx \cdot \partial_z \hbG(\bq,z;z^{\prime}) \bsigtot\cdot{\hat{\bf n}}.
\end{equation}
\end{linenomath*}
Using the Dirac delta relation $\int \exp(-\ii \bq \cdot{\bs}) ds^2 = 4\pi^2 \delta(\bq)$,
\begin{linenomath*}
\begin{equation}
  \label{eq:sig}
 \asw = 4\pi^2 n  \eta \oint_{r=a} \lim_{q\rightarrow 0} \hx \cdot \left[\partial_z \hbG(q,z;z^{\prime})\right]_{z=0}  \bsigtot \cdot{\hat{\bf n}} d S^{\prime}
\end{equation}
\end{linenomath*}
Taking the $q\rightarrow 0$ limit in the full expression for $\partial_z \hbG(q,z,z^{\prime})$ at $z=0$
(see Ref. \cite{Felderhof2005}) leads to a particularly simple expression. For the relevant $xx$ component,
\begin{linenomath*}
  $$\lim_{q\rightarrow 0} \partial_z \hG_{xx}(q,z=0;z^{\prime}) = -\frac{\exp[-\alpha z^{\prime}]}{4\pi \eta}$$
\end{linenomath*}

This allows us to integrate Eq. \ref{eq:sig} and
derive the impedance due to the {\em viscous} stress $\bsig_S$
in \ref{eq:bsigtot},
\begin{linenomath*}
\begin{equation}
  \label{eq:z0}
  \hZ_v(d,R)   \equiv \frac{Z_v}{6\pi n \eta R}  = -\frac{\pi e^{-2\alpha d}}{6}  \left(\frac{2\alpha R \cosh(2\alpha R) -\sinh(2\alpha R)}{\alpha R}\right)
\end{equation}
\end{linenomath*}
and the {\em kinetic} contribution,
\begin{linenomath*}
\begin{equation} 
    \label{eq:z1}
 \hZ_k(d,R)= \\-\frac{2\pi}{3} \frac{e^{-\alpha d}}{\alpha R} \left(\alpha R \cosh(\alpha R) -\sinh(\alpha R)\right) \left(\frac{\rho_p u-\rho \avsx}{\rho \vres}\right).
\end{equation}
\end{linenomath*}

\begin{figure}
 \begin{center}
 \includegraphics[width=0.5\textwidth]{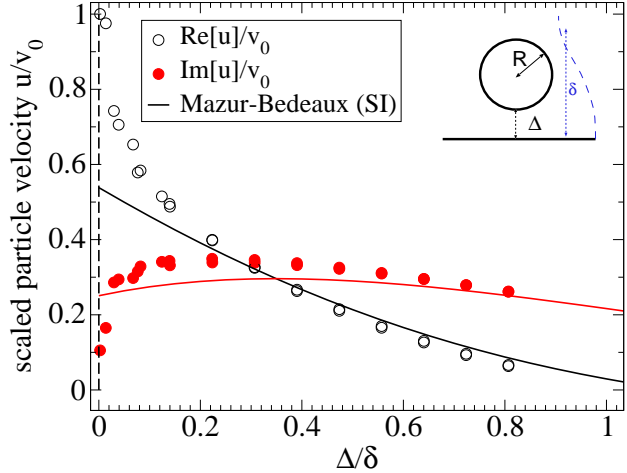}\\
    \end{center}
 \caption{Translational velocity of a spherical particle of radius {\color{black} $R = 50\ \mathrm{nm}$}   suspended over the QCM surface. The distance $\Delta=d-R$ is the gap between the particle surface and the QCM surface, and $\delta$ is the Stokes flow penetration length. Dashed lines correspond to the result of Mazur and Bedeaux relation \cite{Bedeaux1974} (see Supplementary Information) taking the Stokes flow as the mean flow.}
   \label{fig:vel}
\end{figure}

  These expressions apply for a particle suspended at a distance $d$ over the resonator, moving with a velocity $u$ (in turn, $u$
  needs to be determined from the flow-traction, see SI).
  To evaluate the impedance of adsorbed particles $Z^{(ad)}$
  we set  $d=R$ and $u=\vres$ and add the Sauerbrey contribution,
  leading to
  \begin{linenomath*}
  \begin{equation}
    \label{eq:zad}
    \hZ^{(ad)}(R) = -\frac{2 \rho_e}{9\rho} (\alpha R)^2 + \hZ_v(R,R) +\hZ_k(R,R) + \hZ_p \;\; \textrm{    (with } u=\vres\textrm{)}.
  \end{equation}
  \end{linenomath*}
Recall that we neglect the impedance  due to the perturbative flow $\hZ_p$ and later validate such an approximation.  It is interesting to scrutinize the robustness of the ``no-slip'' condition $u=\vres$ to
estimate how feasible it is to get a phase lag between $u$ and $\vres$. To this end Fig. \ref{fig:vel} illustrates the velocity of a
{\em free sphere} moving at a gap-distance $\Delta=d-R$ over
the  oscillating  surface. The case corresponds to $R=50\ \mathrm{nm}$. Solid lines
correspond to the Mazur-Bedeaux relation \cite{Bedeaux1974} (see SI),
which is valid far away from the
surface, as it neglects the reaction field reflected back from the wall.
Notably, even in  the absence  of   wall-particle forces, the  strong hydrodynamic  friction  close to  the  wall  (lubrication) leads to $u\approx     \vres$  as     $\Delta     \rightarrow  0$ (we note that particle slip might take place in specific cases, for instance between two smooth hydrophilic surfaces \cite{2002Bonaccurso}).
If the fluid carries along the particle concomitantly with the wall,
the amplitude of any (distance-dependent) wall-particle force should be small or even  zero, thus creating a small load  impedance.
This fact partially explains why the present theory reproduces so well
a large list of experiments with considerably different colloidal particles and substrates.

In what follows we compare the  prediction in  Eq. \ref{eq:zad} with
3D simulations of spherical rigid particles (Methods) and with 
published experimental data for a wide range of analytes.
We first deal with quasi-neutrally buoyant analytes
(proteins, viruses, liposomes, polymer beads) which possess densities $\rho_p$ that differ from that of the solvent by less than $30\%$
and also treat their mixtures (latex nanoparticles \cite{Olsson2013}).
Secondly, we consider inertial effects in massive particles by comparing the present theory with experiments with
silica nanoparticles in ethanol $\rho_p \approx 2.42\rho$ \cite{Grunewald2015a}.

\subsection{Neutrally buoyant particles}

\subsubsection{Comparison with simulations}

Simulations of neutrally buoyant spheres ($\rho_p=\rho$ in Eq. \ref{eq:z1} and $\rho_e=0$ in Eq. \ref{eq:zad}) were performed using the immersed boundary method combined with an elastic network model for {\color{black}rigid} spheres. Details can be found in \cite{prapp2020} and Methods.
We measured the impedance as a function of the resonator-particle
gap distance $\Delta$ \cite{prapp2020} and here
we consider the limit $\Delta \rightarrow 0$, to deal with the case
of adsorbed particles. Again, it is important to stress that
in these simulations we have not imposed any adhesion force between the particle and
the wall, so their impedance arises only from purely hydrodynamic effects \cite{Melendez2020}.

\begin{figure}
    \begin{center}
 \includegraphics[width=0.6\textwidth]{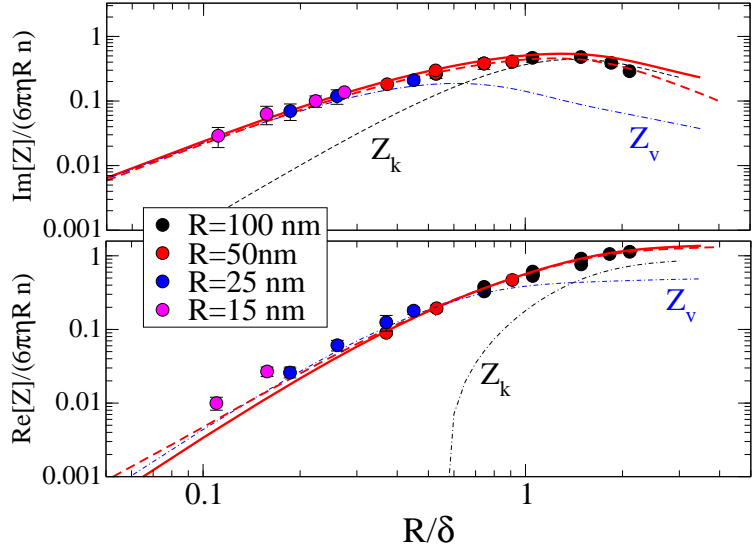}
    \end{center}
    \caption{Scaled  impedances comparing simulations and the
      theoretical result in Eq. \ref{eq:zad} (solid lines). Dashed
      lines correspond to $Z_v$ (Eq. \ref{eq:z0}) and $Z_k$ (\ref{eq:z1}), as indicated.}
    \label{fig:zcomp}
  \end{figure}

Figure  \ref{fig:zcomp} compares  the prediction  in Eq.  \ref{eq:zad}
with simulation  results.  The  agreement is  excellent, both  for the
real and the  imaginary parts of $Z$.  Figure \ref{fig:zcomp}
shows the  contributions to the  hydrodynamic  impedance
in  Eq.  \ref{eq:zad}. The  viscous contribution  $Z_{v}$ dominates
the impedance of small  particles   $R/\delta<0.5$.
Contrary to the commonly assumed relation between
viscous forces and {\color{black}dissipation,} $Z_v$ determines both $\Delta f$ and $\Delta D$
for small $R$. In turn,
for large  particles $R \gtrsim \delta$,
the  inertia of the displaced  fluid $Z_k$ becomes
dominant (although $\mathrm{Re}[Z_v]$ remains
significant).
A {\color{black}maximum}  of $\mathrm{Im}[Z]$ is found {\color{black}near} $R/\delta  \approx 1.5$,  which corresponds to the most negative value of $\Delta f$.
For $R/\delta>2$  (not  shown)  a  sudden  transition  to
$\mathrm{Im}[Z]<0$      ($\Delta f>0$) is expected \cite{Melendez2020,Olsson2011,Tarnapolsky2018}). Interestingly, Eq.  \ref{eq:zad} predicts the cross-over,
but just for any non-zero gap $\Delta >0$. This suggests that the frequency
inversion could be consequence of the counter-flow created by the near-field perturbative current and possibly some particle  velocity  phase-lag  induced   by  slip  or
rotation about the linker point \cite{Tarnapolsky2018}.
The analysis  of this  range of  $R/\delta$ is  left for  a
future contribution.

\begin{table*}
  \caption{List of experiments analyzed in this work, including
    the material (liposomes, viruses, proteins...), the radius of the
    particle and the reported frequency range.
    Soft$^{*}$  liposomes (DMPC at $32^o\mathrm{C}$)
    deform upon adsorption and their diameter in bulk fluid is larger than the height $h$ they expose over the surface.  The given value is $h/2$, rather than the liposome radius in bulk
    (which is about 60 nm \cite{reviakine2012}). The substrate
    b-SLB means biotinylated supported lipid bilayers.}
  \begin{tabular}{l|c|c|c|l}
  \hline
  Material & Particle radius [nm] & Frequency range [MHz] & Label & Reference\\  
  \hline 
Rigid liposome  DPPC & $57\pm 4$ &  $[15-75]$ & DPPC-57 2009 & \cite{Tellechea2009} \\
Rigid liposome  DPPC & $41\pm 2$  &  $[15-75]$ & DPPC-41 2009 & \cite{Tellechea2009} \\
  Cow Pea Mosaic Virus (CPMV) & $14$ & $[15-75]$ & CPMV-14 2009 &   \cite{Tellechea2009} \\
  Rigid liposome  DPPC at $25^o C$ & $41.5$  & $[15-75]$ & DPPC-41 2012 & \cite{reviakine2012} \\
Rigid liposome  DMPC at $10^o C$ & $45$   & $[15-55]$ & DMPC-45 2012 & \cite{reviakine2012}\\
Soft$^{*}$  liposome  DMPC at $32^o C$ & $33$ & $[15-55]$ & DMPC-33 2012 & \cite{reviakine2012} \\
Supported Unimelar
Vesicles & $12.5 \pm 2.5$  & $15, 45$ & b-SUV 2008 & \cite{Bingen2008} \\
Avidin  b-SLB & $2.5\pm 0.5$  & $45$ & Av-SLB 2008 & \cite{Bingen2008} \\
Streptavidin on b-SLB & $2.5\pm 0.5$  & $45$ & Sav-SLB 2008& \cite{Bingen2008} \\
Avidin on b-SLB & $2.5\pm 0.5$  & $35$ & Av-SLB 2010 & \cite{Wolny2010} \\
Streptavidin on b-SLB & $2.5\pm 0.5$  & $35$ & SAv-SLB 2010 & \cite{Wolny2010} \\
Neutravidin on b-SLB & $2.5\pm 0.5$  & $35$ & NAv-SLB 2010 & \cite{Wolny2010} \\
Neutravidin on silica & $2.5\pm 0.5$  & $35$ & NAv-Si 2010 & \cite{Wolny2010} \\
Neutravidin on BSA& $2.5$  & $150$ & NAv 2020 & \cite{aws} \\
Latex NP mixtures & 57 and 12 & 35 & Latex 2013 & \cite{Olsson2013} \\
Polymer NP & 13, 20, 33.5, 70 & 5 & Polymer 2020 & \cite{Sadowska2020} \\
  \hline
\end{tabular}
\label{tab:ar}
\end{table*}

\subsubsection{Comparison with experiments}
 In  QCM, the  frequency is  usually taken  as a
proxy to the surface coverage as in most cases $|\Delta f|$ increases
almost  linearly  with $n$.   {\color{black}However,}  coverage  effects arising  from
hydrodynamic  couplings  between analytes  \cite{Gillissen2017}  often
induce {\color{black}non-monotonic} relations between the dissipation and $n$.  As a
consequence, if  the analyte  size {\color{black}is} typically larger than  proteins   \cite{Milioni2017},
the  acoustic   ratio  $-\Delta D/\Delta f$ decreases with $\Delta f$ \cite{reviakine2012}.
By extrapolating to large $\Delta f$, {\color{black}up to} the {\color{black}intercept}
($\AR=0$), some works \cite{Olsson2013,reviakine2012} found
a way to estimate the particle size by assuming that in such {\color{black}a} limit,
adsorption reaches the close-packed limit, treated as a rigid film via Eq. \ref{eq:sla}.
In many instances the estimated ``Sauerbrey height'' $h$ compares quite well with the
particle diameter \cite{Tellechea2009,reviakine2012,Olsson2013}, but
the procedure was reported to {\color{black}fail severely} in some other cases
(e.g. for massive particles \cite{Grunewald2015a}).

\begin{figure*}
      \begin{center}
 \includegraphics[width=0.4\textwidth,angle=-90]{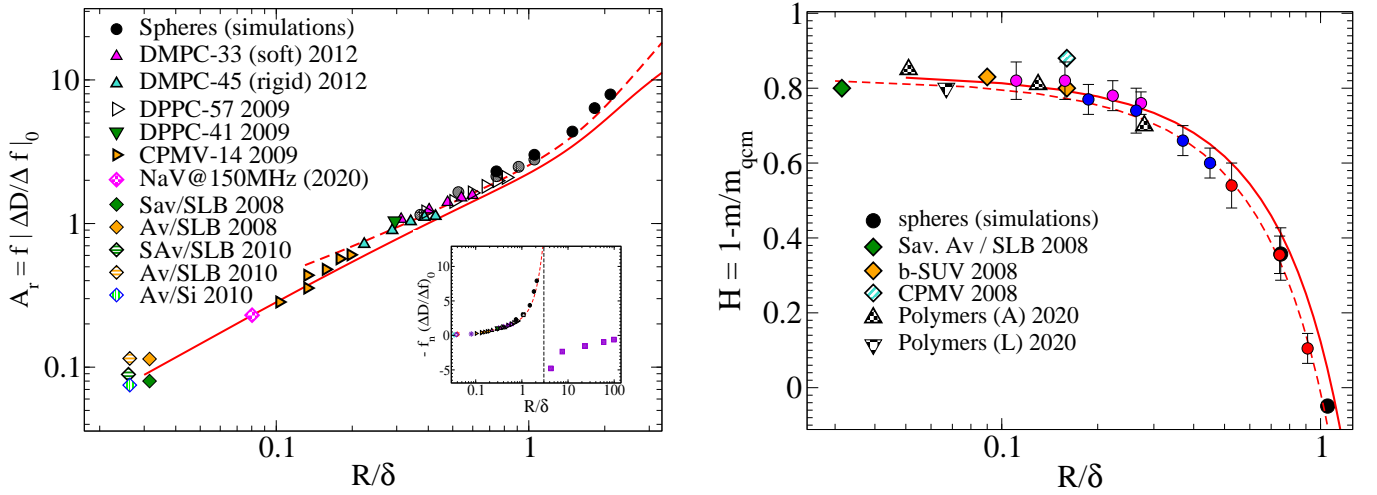}
      \end{center}
      \caption{(a) Non-dimensional acoustic ratio $f|\Delta D/\Delta f|_0$ {\color{black}versus} $R/\delta$
        for experiments  and simulations compared with the theoretical result obtained from Eq. \ref{eq:zad}.
        The solid line is the theoretical trend for $d=R$ (adsorbed particle) and the dashed line corresponds
        to $d=1.05 R$ (small gap between the particle and surface).
        The acronyms used for the legend are explained in Table  \ref{tab:ar}. The inset zooms out the
        particle size range. Beyond the divergence ($R\approx 3 \delta$), the squares correspond to experimental data
        by Sato {\em et al.} \cite{Sato2014} for micron-size particles. (b) Theoretical (lines)
        and experimental evaluation of the mass ratio $H=1-m/\mqcm$ (see text). Filled circles
        {\color{black}correspond} to simulations ({\color{black}color} code similar to Fig. \ref{fig:zcomp}). The label ``CPMV 2008'' in (b)
        indicates data for the cowpea mosaic virus taken from Ref. \cite{Bingen2008}, while in (a) ``CPMV-14 2009''
        is taken from Ref. \cite{Tellechea2009}.}
      \label{fig:ar}
  \end{figure*}

The limit value of the acoustic ratio in the other (dilute) limit $\Delta f\rightarrow  0$,  is
frequently used  to avoid hydrodynamic interactions between analytes (``cross-talk'' effects)
and compare the     ``dissipation     capacity''     of     different     analytes
\cite{Milioni2017,Electra2020,prapp2020}.  This limit acoustic  ratio is taken from the offset $|\Delta D/\Delta  f|_0$ of
the linear fit $-\Delta D/\Delta f = |\Delta D/\Delta f|_0 - a |\Delta f|$.
The present work  focuses on this dilute limit, where particles can be treated
as discrete isolated elements. We deploy the {\em non-dimensional}
acoustic ratio $\ARN\equiv f_n \AR$ which can be extracted from 
the {\color{black}relatively} abundant experimental data.
Figure \ref{fig:ar} shows such comparison between
the prediction of Eq. \ref{eq:zad}
and quite disparate experiments summarized and {\color{black}labeled} in Table \ref{tab:ar}.
Data include proteins, viruses, liposomes and latex particles
ranging from {\color{black}a} few nanometers to {\color{black} a few hundred} nanometers
adsorbed on different substrates. As a first conclusion,
the good agreement with the theory validates our approximation concerning the perturbative stress, at least for $R/\delta \lesssim 2$.
For $R/\delta< 1$ all the data collapses {\color{black}onto} a quasi-linear relation
$\ARN \approx 3 R/\delta$. Interestingly, a linear relation 
(with a smaller prefactor) was also derived from hydrodynamic arguments for the acoustic response of simple fluids {\color{black}to} rough walls in the limit of large corrugation {\color{black}lengths} \cite{Urbakh1994}.
Another point to highlight is the large sensitivity of the
impedance to the gap $\Delta=R-d$ between particle and resonator surfaces.
According to Eq. \ref{eq:zad} a gap as small as $\Delta=0.05 R$
(just 5 nm for a 200 nm particle) creates a measurable
increase in $\ARN$ (see dashed line in Fig. \ref{fig:ar}).
Such sensitivity becomes particularly important as $R/\delta > 2$ because $\Delta f$ gradually vanishes and the acoustic ratio diverges. As shown in the inset of Fig. \ref{fig:ar}, we estimate that the divergence takes place at $R/\delta \sim 3$, which is consistent with the experimental data by Sato {\em et al.} \cite{Sato2014} with micron-sized particles, at the other side of the divergence.
  
The large  disparity of  cases included  in Fig.  \ref{fig:ar} deserve
some   comments.   The   experiments   by  Tellechea   {\em  et   al.}
\cite{Tellechea2009}  correspond to  colloidal particles  on inorganic
surfaces: icosahedral cowpea mosaic {\color{black}viruses}  of 30 nm in diameter (CPMV)
and  extruded dialmitoyl  phophatidyl choline  (DPPC) liposomes,  with
diameters  of 83  nm  (DPPC-41)  and 114  nm  (DPPC-57). These  sizes,
measured  by  dynamic light  scattering  in  bulk, coincide  with  the
Sauerbrey height  $h$ \cite{Tellechea2009} thus confirming  that these
particles do  not deform upon  adsorption (having a well defined size
and spherical  morphology and relatively high  stiffness). Experimental
$\ARN$ for different overtones nicely follow the theoretical
curve.   Reviakine {\textit  et  al.} \cite{reviakine2012}  considered
softer  liposomes which  deform  upon  adsorption on  $\mathrm{TiO_2}$
substrate.    They  used   dimyristoyl  phosphatidyl   choline  (DMPC)
liposomes  of about  90 nm  at temperatures  of $10^o  \mathrm{C}$ and
$32^o \mathrm{C}$,  which are respectively  below and above  the lipid
gel-to-fluid  phase transition  ($T_m\approx  24^o \mathrm{C}$).  DMPC
liposomes are rigid at $10^oC$  while for $T=32^oC$ they substantially
soften  and deform  upon adsorption,  exposing a  height {\color{black}$h\approx  65\ 
\mathrm{nm}$} over  the resonator which is significantly  smaller than
{\color{black}their} diameter in solution.
Despite such deformation, Fig. \ref{fig:ar} shows that the
trend for soft DMPC liposomes
{\color{black}agrees} with our theory if the liposome height $h$ is taken as its effective diameter.
This indicates that the hydrodynamic impedance essentially depends
on how far {\color{black}from the resonator the mass is distributed}
(especially, if the particle inertial mass is zero).

The case of  {\em proteins} allows us to further  explore the scope of
such {\color{black}a}  claim  and  {\color{black}to}  gauge   the  relevance   of  the   substrate.
Fig.  \ref{fig:ar}   includes  values  of  $\ARN$   for  avidin  (Av),
streptavidin  (SAv) and  neutravidin (Nav)  over biotynilnated  supported
lipid  bilayers (b-SLB)  and silica,  taken from  Bingen {\em  et al.}
\cite{Bingen2008} and  Wolny {\em et al.}  \cite{Wolny2010} (see Table
\ref{tab:ar}). Bingen {\em et al.}  compare two quite similar proteins
(Sav and Av)  whose acoustic response over b-SLB only  {\color{black}differs} in their
dissipation      (SaV      is       sligtly      more      dissipative
\cite{Bingen2008}). Remarkably a  purely hydrodynamic theory correctly
captures  the response  of these  proteins  with a radius of about$2.5\ 
\mathrm{nm}$. Such agreement confirms  that collective modes in fluids
persist          up          to          few-nanometer          scales
\cite{DeFabritiis2006,Delgado-Buscalioni2007} which contradicts
the hypothesis of trapped solvent moving in ``solid-like'' fashion
with  the       analyte \cite{Bingen2008,reviakine2012,Sadowska2020}.
Wolny {\em  et al.} \cite{Wolny2010} studied Av,  SAv and NAv
in b-SVB, gold and silica substrates.   Their data (at 45 MHz) on b-SLB
is consistent  with that of Bingen  {\em et al.}  (at  35 MHz). {\color{black}However,}
drastic differences are revealed on gold  and silica. On gold, SAv and
Av  present an  extremely  small acoustic  ration $\ARN\approx  0.016$
which evidences  that these  proteins tightly  collapse onto  the gold
substrate. As reported by Milioni \cite{Milioni2017} Sav on gold forms
an homogeneous surface with a height  ranging in the atomic scale.  {\color{black}By
contrast,}  SAv presents  an extremely  large acoustic  ratio on  silica
($\ARN\approx  1$) which  evidences that  it  is not adsorbed \cite{Wolny2010},
but  in suspension. According to  Eq.   \ref{eq:zad}  (taking   $u$  from
Mazur-Bedeaux theory \cite{Bedeaux1974}, see SI) $\ARN\approx  1$ corresponds to SaV suspended about 15 nm from the surface. By  contrast, Av  in silica  presents
$\ARN\approx  0.075$, which  is consistent  with the  hydrodynamics of
adsorbed spherical  particles.
The response of NAv presents significant  variations
with $\ARN \le 0.15$  and  $ \leq 0.25$  on gold  and
silica \cite{Wolny2010}. According to our theory, the large values of $\ARN$ reported
indicate adsorption of small clusters of proteins (between   $6$  and  $10$   nm  radius, in agreement
with the estimation made by Wolny {\em et al.} \cite{Wolny2010}).
These authors report the presence of relatively   rigid    small aggregates  of NAv in the  stock solution \cite{Wolny2010,Boujday2008}
and, consistently, they observe that the acoustic response of NAv decreased if they increased the centrifugation time
of freshly thawed aliquots \cite{Wolny2010}. In this vein, 
more recent experiments performed at larger fundamental frequency 150 MHz \cite{aws}
report values of the acoustic ratio of NAv in gold which are in agreement with the hydrodynamic result for
single protein deposition, as indicated in Fig.  \ref{fig:ar}. In conclusion,
our analysis indicates the leading role of hydrodynamics, even in the case of proteins. Deviations from the theoretical
hydrodynamic trend trend should help to decipher strong protein deformations, clustering,
substrate-protein and protein-protein interactions.

A particularly enlightening verification  of such statement is offered
by  the mass  ratio  $H=1-m/\mqcm$ routinely  measured  in many  QCM
studies.
In  terms of  impedances, $H=1-n \rho_d \omega/\mathrm{Im}[Z]$  or   $H=1-(4/9) (R/\delta)^2/\mathrm{Im}[\hZ]$
(recall $\hZ\equiv Z/(6\pi \eta  R n)$).  Figure \ref{fig:ar}(b) shows
that the hydrodynamic theory predicts  the experimental values for $H$
for quite  disparate analytes.  This plot collects experiments spreading over
more  than  one  decade, where $H$ was interpreted using versions
of the trapped solvent model \cite{Bingen2008,reviakine2012,Sadowska2020},
which considers that some water molecules move concomitantly with the analyte.
If so, $H$ should not   depend  on  the  overtone  $n$.
Incidentally,  the  first  experiments  \cite{Bingen2008}  considered
small particles ($R/\delta  < 0.2$) for which $H$  is roughly constant in Fig. \ref{fig:ar}(b).
Small discrepancies for the largest $n$ (recall that $\delta \propto n^{-1/2}$)
were mentioned \cite{reviakine2012} and in some cases reported
(notably, the small variation measured
for b-SUV's \cite{Bingen2008} is accurately predicted by the theory).
Using larger polymer  nanoparticles
Sadowska {\em et al.} \cite{Sadowska2020} observed
somewhat larger variations of $H$ with $n$, yet their data in Fig.   \ref{fig:ar}(b)
also nicely agrees with the hydrodynamic theory.
Grunewald {\em et al.} \cite{Grunewald2015} reported even stronger deviations
when studying  heavy  particles, which we analyze hereafter.
As a remark, the only significant deviation from the hydrodynamic trend corresponds to 
the virus capsid (CPMV in Fig. \ref{fig:ar}), which has a larger $H \approx 0.9$.
However, increasing $\rho_P$ in Eq. \ref{eq:zad} actually yields an even slightly
smaller $H$. If so, such deviation is not due to trapped solvent,
but rather to some other mechanism
(specific molecular interaction of the virus with the substrate and/or some partial slip)
which deserves to be revisited.

  \subsection{Mixtures of latex nanoparticles}
\begin{figure}
  \begin{center}
 \includegraphics[width=0.5\textwidth]{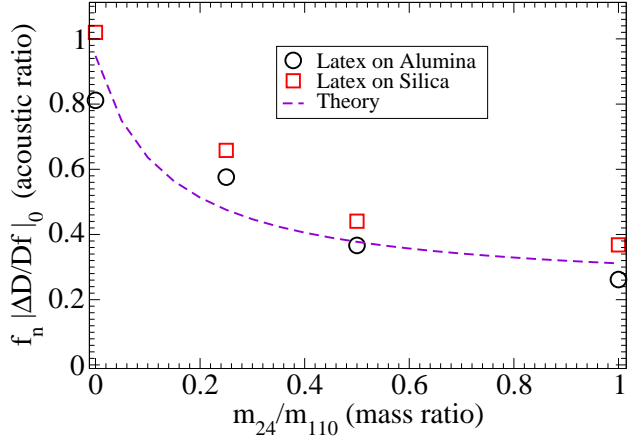}
  \end{center}
  \caption{The acoustic ratio $\ARN = - f (\Delta D/\Delta f)_0$
    for mixtures of latex particles of diameter $24$ and $110 \mathrm{nm}$
    adsorbed {\color{black}on to} Silica or {\color{black}Alumina versus} the mass ratio $m\equiv m_{24}/m_{110}$.
The experimental data was published in Ref. \cite{Olsson2013}.
    The theoretical curve is the weighted impedance (Eq. \ref{eq:zmix})
    $Z_{mix} = \phi Z(R_{24}) + (1-\phi) Z(R_{110})$, where
      the particle number fraction is $\phi = m/(m+r^3)$, with $r$ defined as the diameter-ratio,
      $r\equiv R_{24}/R_{110} =0.218$.}
\label{fig:mix}  
  \end{figure}

  The experiments by Olsson {\em et al.} \cite{Olsson2013}  offers another
  interesting validation of the  present  theory.
  These  authors  considered  mixtures of  latex
  nanoparticles  with   nominal  diameter   of  $24$  and   $110$  nm,
  adsorbed {\color{black}on to} either  {\color{black}silica-} or
  {\color{black}alumina-coated} surfaces. {\color{black}Comparison} between
  the purely hydrodynamic theory and the experiments will illustrate
  to what extent contact forces affect the acoustic response of adsorbed particles.
  The acoustic  ratio against $\Delta f$, reported for $n=3$
  of a 5 MHz AT cut, $f_3=15 \mathrm{MHz}$ permitted us to
  extract values of $\ARN$.  When adding a  mixture of
  nanoparticles, the Sauerbrey-relation \ref{eq:mqcm} offers  an effective
particle size, but it does not provide information on
the mass fraction of the different types of particles (which in the experiment
were known a priori).  In order to apply our theoretical
  result to  these mixtures we  need a  weighted average for  the {\em
    impedance} (note that it is incorrect to average acoustic ratios). The impedance  is proportional to the wall
  stress which  has to be summed up over the total {\em number} of particles.
  We denote $N_{D}$ as the number  of particles  with diameter  $D$ (in  nm).  The
  fraction of  $D=24$ particles is  $\phi=N_{24}/(N_{24}+N_{110})$ and
  using  the simple  relation  $m_{D} \propto  N_{D}  D^3$, we  relate
  $\phi$ with the mass ratio $m\equiv m_{24}/m_{110}$,
  \begin{linenomath*}
  $$
  \phi = \frac{m}{m+r^3},
  $$
  \end{linenomath*}
  where we have defined the ratio-of-diameters as $r=D_{24}/D_{110} \approx 0.218$.
  The weighted average for the impedance is simply,
    \begin{linenomath*}
    \begin{equation}
      \label{eq:zmix}
      Z_{\mathrm{mix}}(m) = \phi(m) Z(D_{24}) + [1-\phi(m)] Z(D_{110}).
    \end{equation}
    \end{linenomath*}
    Theoretical   curves    are   compared    with   experiments
    Fig. \ref{fig:mix}. The agreement is quite good and {\color{black}it} indicates that
    {\color{black}theoretical} approaches can  be used to {\color{black}disentangle}  the fraction of
    nanoparticles size  in a mixture.  In mixtures with more  than two
    components one might use the extra information from $\Delta f$ and
    $\Delta  D$  to  fit  the  mass  fractions  with  the  theoretical
    expressions.  This analysis indicates that contact forces have
    a smaller contribution  than hydrodynamics. Therefore unveiling the
    physical properties of contact forces, wall-induced  physico-chemical interactions or any other molecular feature, first require extracting the leading effect of hydrodynamics from the analysis.

      \subsection{Massive particles: inertia effect}

\begin{figure}
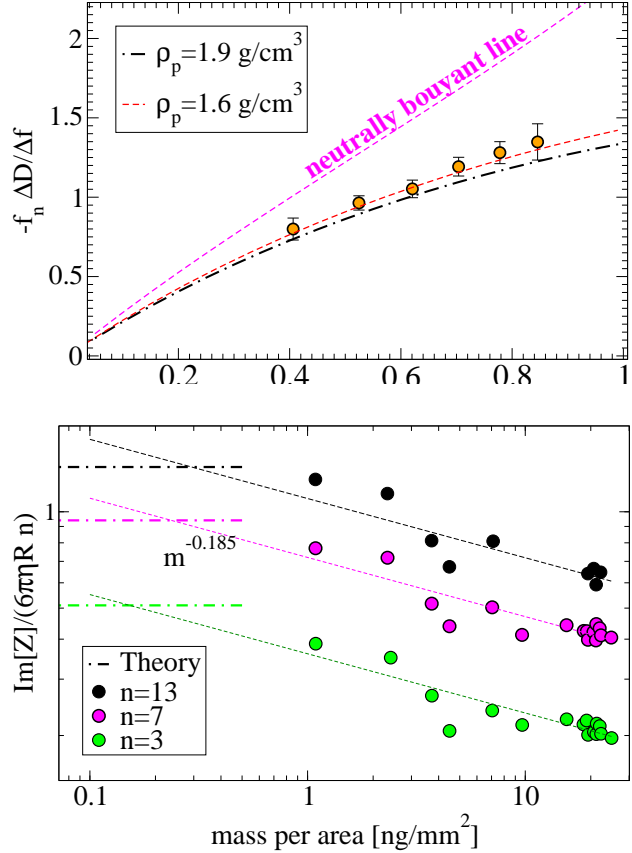

  \begin{center}
\includegraphics[width=0.5\textwidth]{AR_GRUNEWALD.eps}\\
 \includegraphics[width=0.5\textwidth]{FIG2-GRUNEWALD-DFTOT.eps}
  \end{center}
  \caption{Comparison between the theory in Eq. \ref{eq:zad} and 
    experimental data reported by Grunewald {\em et al.} \cite{Grunewald2015a}
    for adsorbed Silica particles in ethanol.
    (a) The acoustic ratio at the low coverage limit ($\Delta f \approx 0$) and (b)
    the scaled impedance versus the adsorbed mass (for several overtones $n$).
    The horizontal dashed lines correspond to the theoretical prediction
    (using the experimental nanoparticle density $\rho_p=1.9\ \mathrm{g/cm^3}$), which is valid for very small coverage.
  }
\label{fig:inertia}  
  \end{figure}

      The     experiments     of     Grunewald     \textit{et     al.}
      \cite{Grunewald2015a} allow  us to  validate our  theory against
      the effect  of particle inertia.  These  experiments studied the
      acoustic   response  of   amine  functionalized   porous  silica
      nanoparticles  strongly   adsorbed  on  gold   surfaces.   These
      nanoparticles,  with  nominal  radius {\color{black}$68.5\  \mathrm{nm}$},  were
      immersed    in    ethanol    at    $T=25^o    C$    ({\color{black}$\rho=0.785\ 
      \mathrm{g/cm^3}$}),  and were  prepared to  present a  repulsive
      electrostatic interaction  which induced an  ordered deposition,
      reaching  a maximum  coverage of  about $15\%$.   Values of  the
      frequency and  dissipation shifts were  obtained for a  range of
      overtones  $n\in  [3,13]$.  The  kinetic  viscosity  of  ethanol
      {\color{black}$\nu=1.33 \times  10^{-6}\ \mathrm{m^2/s}$} yields  a penetration
      length {\color{black}$\delta_n = 292\,  n^{-1/2}\ \mathrm{nm}$} for the $n^{th}$
      overtone    (the   fundamental    resonator   frequency    being
      {\color{black}$f_1=4.95$ MHz}).  Mesoporous  silica nanoparticles  were reported
      to have a void fraction of  about $15 \%$ which yields a density
      in  ethanol of  about {\color{black}  $\rho_p\approx 1.9\ \mathrm{g/cm^3}$}.   The
      authors evaluated  the deposited  mass $\mqcm$  using the
      Sauerbrey   relation  \ref{eq:mqcm},   which   resulted  to   be
      significantly  larger   than  the  deposited  mass   $m$
      evaluated from  the dried sample, using  SEM. Moreover, contrary
      to the  trapped-fluid model  \cite{Bingen2008,Sadowska2020}, the
      mass ratio $H=1- m/\mqcm$ was seen to significantly
      vary with $n$.  We start by comparing our theoretical prediction
      for   the  {\color{black}limiting} acoustic   ratio  $\ARN$, obtained from a linear
      extrapolation of the experimental  data for $-\Delta D/\Delta f$
      to  $\Delta f=0$.  Figure \ref{fig:inertia}  shows an  excellent
      agreement for  the complete  overtone range. Albeit, we noticed
      that  the predicted  $\ARN$  obtained  by inserting  $\rho_p=1.9
      \mathrm{g/cm^3}$  in Eq.  \ref{eq:zad}  slightly underestimates the
      experimental trend.  Incidentally, we found a  better agreement
      using {\color{black}$\rho_p=1.6\ \mathrm{g/cm^3}$} (see Fig .\ref{fig:inertia}).
      However, the  analysis of the experimental  frequency $\Delta  f$
      revealed an interesting surprise: $\Delta f$ increases sublinearly
      with the deposited mass $m$. This fact is revealed
      in  Fig. \ref{fig:inertia}(b): in terms of {\color{black}the scaled} impedance
      $\mathrm{Im}[\hZ]  \sim m^{-0.18(5)}$,  which implies
      $\Delta  f\sim m^{0.81(5)}$. Theoretical predictions for $\mathrm{Im}[\hZ]$
      [using     {\color{black}$\rho_p=1.9\  \mathrm{g/cm^3}$}, plotted as horizontal         lines          in
        Fig.  \ref{fig:inertia}(b)]  consistently  extrapolate  the
      experimental values to  the ultra-dilute  regime   $m  \approx 0.2 \mathrm{ng/mm^2}$ which is close {\color{black}to} or
      below the QCM's limit of detection. In such {\color{black}a} limit,
      $\Delta  f$ becomes slightly larger, which explains the theoretical
      underestimation of $\ARN$  in Fig.  \ref{fig:inertia}(a).   In
      passing,  we   note  that  the sublinear scaling   $\Delta  f\sim m^{0.815}$ is most probably due to hydrodynamic  interaction {\color{black}among}
      silica  particles, but this issue is beyond the present 
      contribution.
      

\section{Discussion}
In summary, the present analytical study on the QCM response of discrete adsorbates shows
that the main source of acoustic impedance comes from the 
hydrodynamic propagation of fluid-induced forces on the analyte. The sensed extra wall stress
strongly depends on how mass is distributed over the resonator. And, in turn, such distribution is determined
by physico-chemical forces (adhesion, dispersion and electrostatic forces, structural elasticity, etc.).
This fact already permits the extraction of relevant information on the underlying microscopic configurations,
uniquely invoking fluid-induced response (as done in the present work).
However, physico-chemical forces are {\em also}
transferred to the fluid and {\em hydrodynamically propagate} to the surface.
An extension of the present theory including these secondary forces (the very purpose of QCM research),
will allow deciphering and {\em measuring} subtle molecular properties, such as the
different acoustic response of avidin and streptavidin, the bending rigidity and membrane fluidity
of liposomes, or the reason behind the deviation from the purely hydrodynamic trend of
the acoustic response of adsorbed virus capsids. 


  \section{Methods}
We have performed {\color{black}three-dimensional}
simulations of the QCM response of elastic spheres {\color{black}with our own} software for Graphical Processors Units
{\tt FLUAM} \cite{fluam,usabiaga2014inertial,Usabiaga2013}
It uses the immersed boundary method (IBM) to couple the hydrodynamics of compressible flows with the dynamics of immersed molecular structures. The integration scheme is second-order accurate in space and time and the spatial discretization is based on a staggered grid \cite{balboa2012staggered} of cell size $h=3.958\ \mathrm{nm}$. Simulations were performed in boxes periodic in the resonator plane.
Boundary conditions for the top and bottom walls were imposed using
a ghost cell to easily impose a tangential velocity {\color{black}
$v_0\cos(\omega t)$ (along the $x$ direction)} at the bottom wall \cite{prapp2020}.
The tangential velocity gradient at the wall $(\partial v_x/\partial z)_{z=0}$
was calculated using a second order spatial interpolation from the upper fluid
cells. The fluid traction (stress) at the resonator is measured by
averaging $\eta (\partial v_x/\partial y)_{y=0}$ over all the surface.
Using the small load approximation, the complex Fourier amplitude
of the average {\color{black}stress} directly leads to the impedance.
Hollow spheres over the resonator (representing liposomes) were modelled using the elastic network model (ENM). The sphere's surface is created by an
arrangement of IBM markers in close packing, connected to their nearest neighbours (at distance $\ell\approx 2h$) by strong harmonic springs.
The bending rigidity of the
structure corresponds to the rigid limit ($k_L \ell^2 \sim 10^5 k_BT$ for $T=300\ \mathrm{K})$). The number of beads required to build the hollow sphere increases as $(R/h)^2$ being about $6000$ beads for a liposome of radius $R=50\, \mathrm{nm}$. 
\section{Acknowledgments}
This work was funded by the EU FET-Open Project ``CATCH-U-DNA''.

\section{Author contributions}
R.D-B and M.M.S conceived the idea, M.M.S. derived a preliminary theoretical approach, R.D-B derived the final equations,
wrote the paper and analyzed the experimental data.

\section{Competing financial interests:} The authors declare no competing financial interests.

\bibliography{qcm4}

\end{document}